# Incoherent boundary conditions and metastates

**Aernout C. D. van Enter**[1], **Karel Netočný**[2] **and Hendrikjan G. Schaap**[1]

*University of Groningen, Academy of Sciences of the Czech Republic*

**Abstract.** In this contribution we discuss the role which incoherent boundary conditions can play in the study of phase transitions. This is a question of particular relevance for the analysis of disordered systems, and in particular of spin glasses. For the moment our mathematical results only apply to ferromagnetic models which have an exact symmetry between low-temperature phases. We give a survey of these results and discuss possibilities to extend them to some situations where many pure states can coexist. An idea of the proofs as well as the reformulation of our results in the language of Newman-Stein metastates are also presented.

## 1. Introduction

In the theory of Edwards-Anderson (short-range, independent-bond Ising-spin) spin-glass models, a long-running controversy exists about the nature of the spin-glass phase, and in particular about the possibility of infinitely pure states coexisting. Whereas on the one hand there is a school which, inspired by Parisi's [22, 24] famous (and now rigorously justified [34, 35]) solution of the Sherrington-Kirkpatrick equivalent-neighbour model, predicts that infinitely many pure states (= extremal Gibbs measures, we will use both terms interchangably) can coexist, the other extreme, the droplet model of Fisher and Huse, predicts that only two pure states can exist at low temperature, in any dimension [18]. An intermediate, and mathematically more responsible, position was developed by Newman and Stein, who have analyzed a number of properties which a situation with infinitely many pure states should imply [25, 26, 27, 28, 29, 30]. One aspect which is of particular relevance for interpreting numerical work is the fact that the commonly used periodic or antiperiodic boundary conditions might prefer different pairs of pure states. Which one then would depend on the disorder realization and on the (realization-dependent) volume, a scenario described by them as "chaotic pairs" (see also [7]). This is a particular example of their notion of "chaotic size-dependence", the phenomenon that the Gibbs measures on an increasing sequence of volumes may fail to converge almost surely (in the weak topology) to a thermodynamic limit. Such a situation may occur when the boundary conditions are not biased towards a particular phase (or a particular set of phases), that is they are not *coherent*. If such pointwise convergence fails, a weaker type of probabilistic convergence, e.g. convergence in distribution, may still be possible. Such a convergence has as its limit objects

[1]Centre for Theoretical Physics, Rijksuniversiteit Groningen, Nijenborgh 4, 9747 AG Groningen, The Netherlands, e-mail: aenter@phys.rug.nl; H.G.Schaap@phys.rug.nl
[2]Institute of Physics, Academy of Sciences of the Czech Republic, Na Slovance 2, 182 21 Prague 8, Czech Republic, e-mail: netocny@fzu.cz

"metastates", distributions on the set of all possible Gibbs measures. (These objects are measures on all possible Gibbs measures, including the non-extremal ones. The metastate approach should be distinguished from the more commonly known fact that all Gibbs measures are convex combinations -mixtures- of extremal Gibbs measures.) For a recent description of the theory of metastates see also [2, 3]. Although our understanding of spin glasses is still not sufficient to have many specific results, the metastate theory has been worked out for a number of models, mostly of mean-field type [4, 5, 11, 19, 20]. Recently we have developed it for the simple case of the ferromagnetic Ising model with random boundary conditions [9, 10, 33]. This analysis is in contrast to most studies of phase transitions of lattice models which consider special boundary conditions, such as pure, free or (anti-)periodic ones.

In this paper we review these results as well as discuss some possible implications for more complex and hopefully more realistic situations of disordered systems, compare [13]. We feel especially encouraged to do so by the recent advice from a prominent theoretical physicist that: " Nitpickers ... should be encouraged in this field.... [12]".

## 2. Notation and background

For general background on the theory of Gibbs measures we refer to [8, 16]. We will here always consider Ising spin models, living on a finite-dimensional lattice $\mathbb{Z}^d$. The spins will take the values 1 and $-1$ and we will use small Greek letters $\sigma, \eta, \ldots$ to denote spin configurations in finite or infinite sets of sites. The nearest-neighbour Hamiltonians in a finite volume $\Lambda \subset \mathbb{Z}^d$ will be given by

$$H^\Lambda(\sigma, \eta) = \sum_{\langle i,j \rangle \subset \Lambda} J(i,j)\, \sigma_i \sigma_j + \sum_{\substack{\langle i,j \rangle \\ i \in \Lambda, j \in \Lambda^c}} J'(i,j)\, \sigma_i \eta_j. \tag{1}$$

Fixing the boundary condition which is denoted by the $\eta$-variable, these are functions on the configuration spaces $\Omega^\Lambda = \{-1, 1\}^\Lambda$. We allow sometimes that the boundary bonds $J'$ take a different value (or are drawn from a different distribution) from the bulk bonds $J$. Our results are based on considering the ferromagnetic situation with random boundary conditions, where the $J$ and the $J'$ are constants ($J < 0$ for ferromagnets), and the disorder is only in the $\eta$-variables which are chosen to be symmetric i.i.d. Associated to these random Hamiltonians $H^\Lambda(\sigma, \eta)$ are random Gibbs measures

$$\mu_\eta^\Lambda(\sigma) = \frac{1}{Z^\Lambda(\eta)} \exp[-H^\Lambda(\sigma, \eta)]. \tag{2}$$

We analyze the limit behaviour of such random Gibbs measures at low temperatures (i.e. $|J| \gg 1$), in the case of dimension at least two, so that the set of extremal infinite-volume Gibbs measures contains more than one element. It is well known that in two dimensions there exist exactly two pure states, the plus state $\mu^+$ and the minus state $\mu^-$. In more than two dimensions also translationally non-invariant Gibbs states (e.g. Dobrushin interfaces) exist. Intuitively, one might expect that a random boundary condition favours a randomly chosen pure translation-invariant state. Such behaviour in fact is expected to hold in considerable generality, including the case where an infinite number of "similar" extremal Gibbs states coexist.

In our ferromagnetic examples all these pure states are related by a (e.g. spin-flip) symmetry of the interaction and the distribution of the boundary condition is also invariant under this symmetry. In more general cases one might have "homogeneous" pure phases not related by a symmetry. In such a case one might need to consider non-symmetric, and possibly even volume-dependent distributions for the boundary conditions to obtain chaotic size-dependence, but for not specially chosen distributions one expects to obtain a single one of these pure states. We will consider here only situations where a spin flip symmetry is present, at least at the level of the disorder distribution. Note further, that in some of these more general situations, in particular situations with many pure states, other types of boundary conditions (e.g. the periodic and the free ones, in contrast to what we are used to) may be not coherent, and they would pick out a random -chaotic- pair of Gibbs states, linked by the spin-flip symmetry.

The physical intuition for the prevalence of the boundary conditions picking out an extremal Gibbs measure is to some extent supported by the result of [15], stating that for any Gibbs measure $\mu$ –pure or not–, $\mu$-almost all boundary conditions will give rise to a pure state. However, choosing some prescribed Gibbs measure to weigh the boundary conditions has some built-in bias, and a fairer question would be to ask what happens if the boundary conditions are symmetric random i.i.d. We will see that the above intuition, that then both interfaces and mixtures are being suppressed, is essentially correct, but that the precise statements are somewhat weaker than one might naively expect.

As a side remark we mention that biased, e.g. asymmetric, random boundary conditions are known to prefer pure states. This is discussed in [6, 17] for some examples. A similar behaviour pertains for boundary conditions interpolating between pure and free, as is for example shown for some particular cases in [1, 21].

## 3. Results on Ising models with random boundary conditions

In this section we describe our results on the chaotic size behaviour of the Ising model under random boundary conditions in more detail. We consider the sequences of finite-volume Gibbs states $\mu_\eta^\Lambda$ along a sequence of concentric cubes $\Lambda_N$ with linear size $N$, for any configuration $\eta \in \Omega = \{-1,1\}^{\mathbb{Z}^d}$ sampled from the symmetric i.i.d. distribution with the marginals $\text{Prob}(\eta_i = -1) = \text{Prob}(\eta_i = 1) = \frac{1}{2}$. To any such sequence of states we assign the collection of its weak-topology limit points, which can in general be non-trivial and $\eta$-dependent. However, in our simple situation we show that the set of limit points has a simple structure: with probability 1, it contains exactly two elements – the Ising pure states $\mu^+$ and $\mu^-$. This is proven to be true, provided that a sufficiently sparse (depending on the dimension) sequence of cubes is taken and for sufficiently low temperatures ($-J \gg 1$), at least in certain regions in the $(J, J')$-plane which will next be described. Although one gets an identical picture in all the cases under consideration, these substantially differ in the complexity of the analysis required in the proof.

### *Ground state with finite-temperature boundary conditions*

As a warm-up problem, following [9], we consider the case where $-J = \infty$ and $J'$ is finite. Then all spins inside $\Lambda$ take the same spin value, either plus or minus. Our choice of coupling parameters has excluded interface configurations. The total energy for either the plus or the minus configuration of the system in a cube $\Lambda_N$,

$$H^N(\pm, \eta) = \pm J' \sum_{i:\, d(i,\Lambda_N)=1} \eta_i \tag{3}$$

is a sum of $\mathcal{O}(N^{d-1})$ 2-valued random variables, which are i.i.d. and of zero mean. Obviously,

$$\mu_\eta^N(\pm) = \left\{1 + \exp[\pm 2 H^N(+, \eta)]\right\}^{-1} \tag{4}$$

and the possible limit states are the plus configuration, the minus configuration, or a statistical mixture of the two, depending on the limit behaviour of the energy $H^N(+, \eta)$. According to the local limit theorem, the probability of this energy being in some finite interval decays as $N^{-\frac{d-1}{2}}$. Summing this over $N$ gives a finite answer if either $d$ is at least 4, or if one chooses a sufficiently sparse sequence of increasing volumes $\Lambda_{N_k}$ in $d = 2$ or $d = 3$. A Borel-Cantelli argument then implies that almost surely, that is for almost all boundary conditions, the only possible limit points are the plus configuration and the minus configuration. On the other hand, without the sparsity assumption on the growing volumes in dimension 2 and 3, again via a Borel-Cantelli argument, a countably infinite number of statistical mixtures is seen to occur as limit points. However, as these mixtures occur with decreasing probabilities when the volume increases, the metastate of our system is concentrated only on the plus and minus configuration, and the mixtures do not show up (they are null-recurrent). We expect a similar distinction between almost sure and metastate behaviour in various other situations.

### *Finite low temperatures with weak boundary conditions*

In the case where $J$ and $J'$ are both finite, but $|J'| \ll -J$, the spins inside $\Lambda$ are no more frozen and thermal fluctuations have to be taken into account. Yet, one can expect that for the bulk bond $-J$ being large enough, the behaviour does not change dramatically and the model can be analyzed as a small perturbation around the $-J = \infty$ model of the last section. A difference will be that the frozen plus and minus configurations are to be replaced with suitable plus and minus ensembles, and the energies with the free energies of these ensembles. A technical complication is that these free energies can no longer be written as sums of independent terms. This prevents us from having a precise local limit theorem. However, physically the situation should be rather similar, and one can indeed prove a related, but weaker result.

To describe the perturbation method in more detail, we need to go to a contour description. Every pair of spin configurations related by a spin-flip corresponds to a contour configuration. We will distinguish the ensemble of plus-configurations $\Omega_+^\Lambda$ in which the spins outside of the exterior contours are plus and similarly the ensemble of minus-configurations $\Omega_-^\Lambda$. When a contour ends at the boundary and separates one corner from the rest, this corner is defined to be in the interior, and for contours separating at least two corners from at least two other corners (these are interfaces of some sort) we can make a consistent choice for what we call the interior, see [9, 10]. It will turn out that our results do not depend on our precise choice. We consider the measures restricted to the plus and minus ensembles:

$$\mu_{\eta,+}^\Lambda(\sigma) = \frac{1}{Z^\Lambda(\eta, +)} \exp[-H^\Lambda(\sigma, \eta)] \, \mathbf{1}_{[\sigma \in \Omega_+^\Lambda]} \tag{5}$$

and similarly for the minus ensemble. Under the weakness assumption $|J'| \ll -J$, the Gibbs probability of any interface in these ensembles is damped exponentially in the system size $N$ uniformly for all $\eta$. Actually, we can prove the asymptotic triviality of both ensembles in the following strong form:

**Proposition 3.1.** *Let* $-J \geq \max\{J^*, \Delta|J'|\}$ *with large enough constants* $J^*, \Delta > 0$. *Then* $\mu_{\eta,\pm}^{\Lambda_N} \xrightarrow{N} \mu^{\pm}$ *in the weak topology. (The convergence is exponentially fast uniformly in* $\eta$.)

**Remark.** Note that for Dobrushin boundary conditions, if $\tilde{\Delta}|J'| > -J$ for some sufficiently small $\tilde{\Delta}$, these boundary conditions favour an interface. However, such boundary conditions are exceptional, and in the next section we will prove a weaker statement, based on this fact, in $d = 2$.

The convergence properties of the full, non-restricted measures are based on an estimate for the random free energies:

$$F_{\pm}^{\Lambda}(\eta) = \log Z^{\Lambda}(\eta, \pm) \tag{6}$$

namely, we prove the following weak local limit type upper bound:

**Proposition 3.2.** *Under the assumptions of Proposition 3.1, the inequality*

$$\mathrm{Prob}(|F_+^{\Lambda_N}(\eta) - F_-^{\Lambda_N}(\eta)| \leq N^{\varepsilon}) \leq \mathrm{const}\, N^{-(\frac{d-1}{2}-\varepsilon)}$$

*holds for any* $\varepsilon > 0$ *and* $N$ *large enough.*

The proofs of both propositions are based on convergent cluster expansions for the measures $\mu_{\eta,+}^{\Lambda}$ and the characteristic function of the random free energy difference, respectively. Combining Proposition 3.2 with a Borel-Cantelli argument we get

$$\lim_{\Lambda} |F_+^{\Lambda}(\eta) - F_-^{\Lambda}(\eta)| = \infty \tag{7}$$

provided that the limit is taken along a sparse enough sequence of cubes (unless $d \geq 4$). By symmetry, this reads that the random free energy difference has $+\infty$ and $-\infty$ as the only limit points. Since the full Gibbs state is a convex combination of the plus and the minus ensembles with the weights related to the random free energy difference,

$$\mu_{\eta}^{\Lambda} = \left[1 + \frac{\mathcal{Z}_{\eta,-}^{\Lambda}}{\mathcal{Z}_{\eta,+}^{\Lambda}}\right]^{-1} \mu_{\eta,+}^{\Lambda} + \left[1 + \frac{\mathcal{Z}_{\eta,+}^{\Lambda}}{\mathcal{Z}_{\eta,-}^{\Lambda}}\right]^{-1} \mu_{\eta,-}^{\Lambda} \tag{8}$$

this immediately yields the spectrum of the limit Gibbs states [9].

### *Finite low temperatures in $d = 2$, with strong boundary conditions*

In the case $-J = |J'|$, due to exceptional (e.g. Dobrushin-like, all spins left minus, all spins right plus) boundary conditions, we are to expect no uniform control anymore over the convergence of the cluster expansion. Indeed, one checks that for the contours touching the boundary the uniform lower bounds on their energies cease holding true. In order to control the contributions from these contours, we need to perform a multiscale analysis, along the lines of [14], with some large deviation estimates on the probability of these exceptional boundary conditions. We obtain in this way the following Propositions, corresponding to Propositions 3.1-3.2 [10]:

**Proposition 3.3.** *Let* $d = 2$ *and* $|J'| = -J \geq J^{**}$, *with the constant* $J^{**} > 0$ *being large enough. Then* $\mu_{\eta,\pm}^{\Lambda_N} \xrightarrow{N} \mu^{\pm}$ *weakly for almost all* $\eta$.

**Proposition 3.4.** *Under the same assumptions as in Proposition 3.3,*

$$\text{Prob}(|F_+^\Lambda(\eta) - F_-^\Lambda(\eta)| \leq \tau) \leq \text{const}(\tau)\, N^{-(\frac{1}{2}-\varepsilon)} \qquad (9)$$

*for any $\tau, \varepsilon > 0$ and $N$ large enough.*

Observe that the limit statement of Proposition 3.3 holds true only on a probability 1 set of the boundary conditions (essentially the ones not favouring interfaces). The construction of this set is based on a Borel-Cantelli argument.

These two Propositions imply the following Theorems on the almost sure behaviour and the metastate behaviour respectively:

**Theorem 3.1.** *Let the conditions of either Proposition 3.1 or Proposition 3.3 be satisfied, and take a sequence of increasing cubic volumes, which in $d = 2$ and $d = 3$ is chosen sufficiently sparse. Then for almost all boundary conditions $\eta$ it holds that the weak limit points of the sequence of finite-volume Gibbs states are the plus and minus Ising states. Almost surely any open set in the set of Gibbs states, from which the extremal Gibbs measures have been removed, will not contain any limit points.*

**Theorem 3.2.** *Let the conditions of either Proposition 3.1 or Proposition 3.3 be satisfied, and take a sequence of increasing cubic volumes. Then the metastate equals the mixture of two delta-distributions: $\frac{1}{2}(\delta_{\mu^+} + \delta_{\mu^-})$.*

**Remark 3.1.** This metastate is a different one from the one obtained with free or periodic boundary conditions which would be $\delta_{\frac{1}{2}(\mu^+ + \mu^-)}$. In simulations it would mean that for a fixed realization and a fixed finite volume one typically sees the same state (either plus or minus, which one depending on the volume), the other one being invisible. For periodic or free boundary conditions both plus and minus states are accessible for any fixed volume.

**Remark 3.2.** We expect that, just as in the ground state situation, for a non-sparse increasing sequence of volumes, mixtures will be null-recurrent in dimensions 2 and 3. In this case, any mixture, not only a countable number of them, could be a null-recurrent limit point. However, as our Propositions only provide upper bounds, and no lower bounds, on the probabilities of small (and not even finite) free energy differences, such a result is out of our mathematical reach.

Note that although the metastate description of Theorem 2 gives less detailed information than the almost sure statement of Theorem 1, it encapsulates the physical intuition actually better.

**Remark 3.3.** It also follows from our arguments that, for almost all boundary conditions $\eta$, large (proportional to $N^{d-1}$) contours are suppressed for large systems, so neither rigid, nor fluctuating interfaces will appear anywhere in the system. Fluctuating interfaces also would produce mixed states, so even in 2 dimensions, where interface states do not occur, this is a non-trivial result.

**Remark 3.4.** Similar results can also be obtained for the case of very strong boundary conditions, $|J'| \gg -J \gg 1$, [33]. Our method can easily be adapted to get the same result for all $|J'| \leq -J$, and we believe in fact the result to be true for all $J' \neq 0, -J \gg 1$.

## 4. Spin glasses, Fisher-White type models, many states, space versus time

Although our original motivation was due to our interest in the spin-glass model, our results for spin-glass models are still very modest. We note, however, that by a gauge transformation the ferromagnet with random boundary conditions is equivalent to a nearest-neighbour Mattis [23] spin-glass (= one-state Hopfield) model with fixed boundary conditions, the behaviour of which we can thus analyze by our methods.

In a recent paper by Fisher and White [13], they discuss possible scenarios of infinitely many states, mostly based on constructions of stacking lower-dimensional Ising models. The motivation is to investigate what might happen in the putative infinite-state spin-glass scenario, for some (high-dimensional and/or long-range) Edwards-Anderson type model.

One of their simplest models is based on "stacking" lower-dimensional Ising models, and we can derive some properties for such models from our results in the last section.

For example one can consider $N$ 2-dimensional Ising models in squares of size $N$ by $N$ in horizontal planes, decoupled in the vertical direction. There are now $2^N$ ground states for free boundary conditions, corresponding to the choice of plus or minus in each plane. To avoid reducing the problem to studying the symmetry group of the interaction, one can randomize the problem some more, as follows: By choosing in every plane the bonds across a single line through the origin (in that plane) randomly plus or minus, one obtains that in every plane in the thermodynamic limit four ground states appear, (plus-plus, plus-minus, minus-plus and minus-minus). By imposing random boundary conditions, independently in each plane, one now expects an independent choice out of these four in each plane. By similar considerations as in the last section, we find that although this is true in most planes, in $O(\sqrt{N})$ of them (that is a fraction of $O(\frac{1}{\sqrt{N}})$) of them) mixtures will appear. If our boundary conditions are finite-temperature, this happens if the difference between the number of pluses and minuses stays bounded, otherwise this happens when there is a tie, and at the same time not such a strong spatial fluctuation that an interface ground state appears. The same remains true if one connects the planes by a sufficiently weak (one-dimensional) random coupling, such as for example having a random choice for all the vertical bonds on lines, each line connecting a pair of adjacent planes.

If one would choose the values of *all* vertical bonds randomly from some symmetric distribution, we would obtain something like a collection of two-dimensional random field Ising models. Indeed, the effect of one plane on the next one would be like that of a random field, which would prevent a phase transition in any plane, and our arguments break down. For further discussion see [13]. Similar properties hold for models in which on one periodic sublattice one stacks in a horizontal direction, and in another one in a vertical direction.

As the topology of weak convergence is a topology of convergence of *local* observables, a statement along the lines of Theorem 1 still holds, however the interpretation that one almost surely avoids mixtures now is incorrect. In fact, in these circumstances, for almost all boundary conditions, the system will be in a *mixed* state. Note that, although such mixed states are less stable than interface states (one can change the weights in the mixture by a finite-energy perturbation), they are much more likely to occur than interface states. As the set of sites which are influenced by the measure being a mixture has a density which approaches zero, in the thermodynamic limit the Parisi overlap distribution [24] will be trivially concentrated at 1.

Similarly to Theorem 2, also the metastate still will be concentrated on the symmetric mixture of the delta-distributions on the pure states in each plane.

We obtain similar results at sufficiently low temperatures, although again, now we cannot prove the occurrence of mixtures.

It has been conjectured that similar to the chaotic size-dependence we have discussed, for random (incoherent) initial conditions a related phenomenon of chaotic time-dependence (non-convergence) might occur in disordered systems. If even the Cesaro average fails to converge, this phenomenon also has been called "historic" behaviour [31].

A similar distinction in the spatial problem also occurs. Although for ferromagnets with random boundary conditions the Cesaro average of the magnetisation at the origin, taken over a sequence of linearly increasing volumes, will exist, one does not expect this for the random field Ising model with free or periodic boundary conditions, essentially for the same reason as in the mean-field version of the problem [19, 20]. One would only expect a convergence in distribution to the metastate $\frac{1}{2}(\delta_{\mu^+} + \delta_{\mu^-})$.

One could both scenarios (chaotic size dependence and chaotic time dependence) describe as examples of what Ruelle [32] calls "messy" behaviour.

However, chaotic time dependence for stochastic dynamics of for example Glauber type requires the system to be infinitely large. Otherwise, for finite systems either one has a finite-state Markov chain with a unique invariant measure (for positive temperature Glauber dynamics), while for zero-temperature dynamics the system would get trapped, up to zero-energy spin flips.

Concluding, we have illustrated how the notion of chaotic size-dependence naturally occurs, already in the simple ferromagnetic Ising model, once one allows for incoherent boundary conditions. As in many situations the choice of incoherent boundary conditions seems a physically realistic one, our results may be helpful in explaining why in general one expects experimentally to observe pure phases.

## Acknowledgments

This research has been supported by FOM-GBE and by the project AVOZ10100520 of the Academy of Sciences of the Czech Republic. We thank Christof Külske for helpful advice on the manuscript. A. C. D. v. E. and K. N. thank Igor Medved' for their related earlier collaboration which led to the results in [9]. A. C. D. v. E. thanks Pierre Picco and Veronique Gayrard for useful dicussions on metastate versus almost sure properties. He would like to thank Mike Keane for all the probabilistic inspiration he has provided to Dutch mathematical physics over the years.